\documentclass[%
 reprint,
 amsmath,amssymb,
 aps,
prl,
]{revtex4-2}

\usepackage{booktabs}   
\usepackage{amssymb}
\usepackage{color}
\usepackage{graphicx}
\usepackage{dcolumn}
\usepackage{bm}

\usepackage{amsmath}
\usepackage[pdftex,plainpages=false,colorlinks=true,linkcolor=blue, citecolor=blue, urlcolor=blue]{hyperref}

\newcommand{\bb}{\bullet}
\newcommand{\cc}{\circ}

\begin{document}

\preprint{APS/123-QED}

\title{Strong-coupling Kohn-Luttinger superconductivity in trilaminar Hubbard model: synthesizing unconventional superconductivity in  atomically thin layered materials
}

\title{Strong-coupling Kohn-Luttinger pairing in multi-component Hubbard model: A Route to Synthesizing Unconventional Superconductivity in Stacked 2D Materials
}

\title{ Strong-Coupling Kohn-Luttinger Pairing in Hubbard models : Engineering an Unconventional Superconductivity  in Trilaminar 2D electrons}

\title{ Strong-Coupling Kohn-Luttinger Pairing in Hubbard models : Engineering an Unconventional Superconductivity  in Trilaminar 2D electrons}

\title{Strong-Coupling Kohn-Luttinger pairing in Trilaminar 2D electrons}

\title{Synthesizing Strong-Coupling Kohn-Luttinger Superconductivity in 2D Van der Waals Materials}


\author{
Shi-Cong Mo\textsuperscript{1},
Hongyi Yu\textsuperscript{2},
W\'ei W\'u\textsuperscript{1,}\footnotemark[1]
}
\email{Corresponding author: wuwei69@mail.sysu.edu.cn}

\affiliation{
\textsuperscript{1} Guangdong Provincial Key Laboratory of Magnetoelectric Physics and Devices; School of Physics, Sun Yat-sen University, Guangzhou 510275, China \\
\textsuperscript{2} Guangdong Provincial Key Laboratory of Quantum Metrology and Sensing, School of Physics and Astronomy, Sun Yat-Sen University (Zhuhai Campus), Zhuhai 519082, China
}

\date{\today}


\begin{abstract}
The Kohn-Luttinger (KL) mechanism of pairing, which describes  superconductivity emergent from repulsive interactions, typically yields Cooper pairs at high angular-momentum ($\ell > 0$) and extremely low transition temperatures ($T_c$). Here, we reveal an inter-layer s-wave ($\ell=0$) KL superconductivity with greatly elevated $T_c$  in a multi-layer  Hubbard model, which prototypes stacked two-dimensional (2D) electrons in layered van der Waals materials. By employing determinant quantum Monte Carlo and dynamical mean-field theory simulations, we show that a strong pairing attraction $V^{*}$, without the mediation of collective modes,  can emerge between inter-layer electrons in the system. As inter-layer repulsion $U$ increases,   $V^{*}$ evolves from a conventional KL relation of $V^{*} \propto -U^2$, to a linear strong-coupling scaling of $V^{*} \propto -U$, resulting in enhanced superconductivity at large $U$. This strong-coupling KL pairing is robust against changes in lattice geometries and dimensionalities, and it can  persist, in the presence of a large remnant Coulomb  repulsion $U^{*}$ between pairing electrons. Using \textit{ab initio} calculations, we propose a few 2D layered  van der Waals materials that can potentially realize and control this unconventional superconductivity. 
\end{abstract}

\maketitle

\textit{Introduction-}
 Van der Waals (VdW) materials hosting two-dimensional electrons have emerged as versatile platforms for engineering and studying many exotic correlated phases of matter, including quantum Hall states~\cite{serlin2020intrinsic,xie2022valley,FanMoTe2}, correlated Mott insulating states ~\cite{cao2018correlated,xu2020correlated,regan2020mott,huang2021correlated},  excitonic insulators~\cite{xienematic, nguyen2025perfect,qi2025perfect}
, and superconductivity~\cite{cao2018unconventional,lu2019superconductors,yankowitz2019tuning,xi2016ising,siegl2025friedel,
xia2025superconductivity,guo2025superconductivity,berg2021,siegl2025friedel}.
This field is rapidly advancing due to the unparalleled electrical tunability, broad material palette, and moiré engineering capabilities of VdW materials~\cite{novoselov20162d,zhai2025twistronics}, which continue to promise the discovery of new quantum phases.

On the other hand, in condensed matter physics, the discovering and understanding of new mechanisms of unconventional  superconductivity
~\cite{bednorz1986,sun2023signatures,kotliar1988,zhang1988,moriya1990antiferromagnetic,emery1995,Lee2006review, norman2011challenge,crepel2021new,
abinitio_htc,christos2023model,crepel2023,kumar2024unconventional,dong2025controll,von2024sc,zerba2024realizing}, \textit{i.e.}, in which electron pairing arises sans an electron-phonon-coupling (EPC)  origin, have been one of the central research topics. The magnetic-correlation mediated superconductivity (SC) that relevant to cuprates~\cite{maier2008dynamics,kyung2009pairing,dong2022quantifying,liu2025interplay}, iron-based~\cite{chubu2008,wang2011electron}, and nickelates~\cite{middey2016physics,nomura2022superconductivity,wu2024superexchange,mo2025intertwined} superconductors marks such an example.
Another prominent theoretical paradigm of unconventional SC is the Kohn-Luttinger (KL) superconductivity in metals, where overscreened Coulomb interactions can induce Cooper pairs at high angular-momenta ($\ell > 0$)~\cite{kohn1965new,kagan2014kohn}. Despite its long-standing theoretical proposal, direct experimental evidence for KL SC remains elusive. Recent studies suggest that  SC in 
 graphene systems~\cite{gonzalez2019kohn,cea2021coulomb,geier2025chiral}, and monolayer $\mathrm{NbSe_2}$ ~\cite{siegl2025friedel} could potentially be explained within the KL framework.

\begin{figure}[t!]
\includegraphics[scale=1.3]{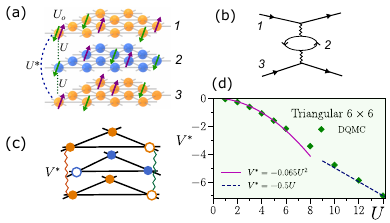}
\caption{Illustration of the trilayer Hubbard model and  KL electron pairing on 2D triangular lattice. \textbf{(a):} AAA-stacking trilayer triangular lattice with spinfull electrons. Different Hubbard terms $U_{l \sigma,m \sigma^{'}}$  within a unit cell are depicted. Long-range interaction $U_{mq}(r)$ terms beyond a unit cell are not shown here.
\textbf{(b):} A KL second order Feynman diagram that leads to  inter-layer  attraction between a pair of electrons in the top- and bottom- layer respectively, via the polorization of the middle-layer. \textbf{(c):} A cartoon showing the inter-layer pairing of two electrons /holes (solid/hollow dots) in  real-space (indicated by wavy line). The spin label is omitted in (b) and (c) for clarity. \textbf{(d):} Effective pairing attraction $V^*$  as a function of $U$ for spinless trilayer Hubbard model from  $6 \times 6 \times 3$ DQMC simulation at $T=0.5, n=0.6$, \textit{cf.} Fig.~\ref{fig:chi}.
}
\label{fig:illu}
\end{figure}

In their original proposal, Kohn and Luttinger demonstrated that electron pairing can occur in metals without involving EPC,  below   $T_c / T_F  \sim exp[-(2\ell)^4]$ at large $\ell$, where $T_F$ is the Fermi temperature~\cite{kohn1965new}.  While such pairing could, in principle, lead to SC, the exponentially suppressed $T_c$ makes experimental observation challenging. Subsequent theoretical studies have suggested that lattice effects~\cite{baranov1992superconductivity,raghu2010superconductivity}, van Hove singularities~\cite{gonzalez2008kohn,nandkishore2012chiral,kagan2014kohn}, the presence of multiple Fermi surfaces~\cite{chubukov2017superconductivity}, and quantum geometry~\cite{shavit2025quantum,jahin} may enhance KL SC. In these studies, KL pairing typically favors a $d-$ wave ($\ell =2$)~\cite{raghu2010superconductivity,nandkishore2012chiral}, or $p-$ wave ($\ell =1$) pairing  symmetry ~\cite{raghu2010superconductivity,kagan2014kohn,geier2025chiral}. 
 There are also  studies
 suggest that for weakly repulsive electrons in a single-band system, pairing with $\ell<3$ is unlikely to occur~\cite{alexandrov2011unconventional}.
  These conclusions are primarily derived from weak-coupling considerations. The behaviour of KL SC at strong-coupling remains unclear to date.

In this work, we propose that trilaminar 2D electron systems in stacked VdW materials (see Fig.~\ref{fig:illu}a), can be utilized to realize an inter-layer $s-$ wave  KL superconductivity. In the strong-coupling regime where inter-layer repulsion $U$  becomes comparable to electronic bandwidth $W$ ($U \sim W$),  we reveal a $V^{*} \propto -U$ scaling  for the effective pairing attraction $V^{*}$. This behavior contrasts with the conventional weak-coupling KL pairing where $V^{*} \propto -U^2/t$, and the magnetic-correlation mediated SC in cuprates where $V^{*} \propto -t^2/U$, $U$ and $t$ are respectively the electron repulsion and hopping amplitudes. We show that such a linearly growing $V^*$ leads to  enhanced
  KL SC at large $U$. We analyse the dependence of pairing on electron densities and
long-range Coulomb repulsions. Using density functional theory (DFT) and constrained random phase approximation (cRPA) calculations, we identify several candidate 2D layered half-metals and metals, including transition-metal halides~\cite{kim2019evolution}, as potential platforms for realizing this strong-coupling KL SC.

\textit{Model and Methods -}
We consider a repulsive Hubbard model on 2D trilayer triangular lattice as following,

\begin{eqnarray}
\mathcal{H}=-t\sum_{\langle i,j\rangle}\sum_{l \sigma}\left(c_{il\sigma}^{\dagger}c_{jl\sigma}^{\phantom{\dagger}}+c_{jl\sigma}^{\dagger}c_{il\sigma}^{\phantom{\dagger}}\right) \nonumber \\ 
+\sum_{il\sigma}(\epsilon_{l}-\mu)n_{il\sigma}  
+\sum_{i}\sum_{l\sigma \neq m\sigma^{\prime}}\frac{U_{l\sigma, m \sigma'}}{2}n_{il\sigma}n_{im\sigma'} \nonumber \\
+\sum_{i,r,l,m,\sigma,\sigma^{\prime}}U_{lm}(r)n_{i,l,\sigma} n_{i+r,m,\sigma'}
\label{eq:hamil}
\end{eqnarray}

where $c^{\dagger}_{il\sigma}$ ($c_{il\sigma}$) denotes the creation (annihilation) operator for  electrons at unit cell $i$, layer $l$, and  with spin flavor $\sigma$.
 The intra-layer nearest-neighbor hopping amplitude $t$ will be used as the
energy unit throughout the paper ($t \equiv1$ ).
The electron density $n_l = \langle n_{l} \rangle$  in each layer  can be tuned by the chemical potential $\mu$ and  site-energies $\epsilon_{l}$.
Unless specified, we maintain identical density for all three layers: $n_1=n_2=n_3 = n$. The Hubbard term $U_{l\sigma, m \sigma^{\prime}}$ describes  repulsions between  two electrons in the same unit cell with  layer and spin indices ($l, \sigma$) and $(m,\sigma^{\prime} )$ respectively,  $(l, \sigma) \neq (m,\sigma^{\prime} )$, 
 $U_{l\sigma, m \sigma^{\prime}}\geq0$ in our study. $U_{lm}(r)$ denotes intra- or inter-layer long-ranged Coulomb repulsions beyond 
 a unit cell.
 Here, we will study both spinless ($N_{\sigma}=1$) and spinful ($N_{\sigma} =2 $) cases ( See Fig.1). We consider three major term of Hubbard $U_{l\sigma, m \sigma^{\prime}}$: repulsion between electrons at neighboring layers $U \equiv U_{1\sigma, 2\sigma'} \equiv U_{2\sigma,3\sigma'}$, and  $U^{*} \equiv U_{1\sigma,3\sigma'}$ for electrons on the top- and bottom-layer. There are also on-site repulsions $U_o \equiv U_{l\uparrow, l\downarrow}$ in the spinfull case. 
 The effects of non-local $U_{lm}(r)$ will first be neglected, and be investigated at the end of the work. To solver Eq.~\ref{eq:hamil}, we employ  determinant quantum Monte Carlo (DQMC)  and dynamical mean-field theory (DMFT)~\cite{georges96} approaches.

\begin{figure}[t!]
\centering
\centering{}\includegraphics[scale=0.65]{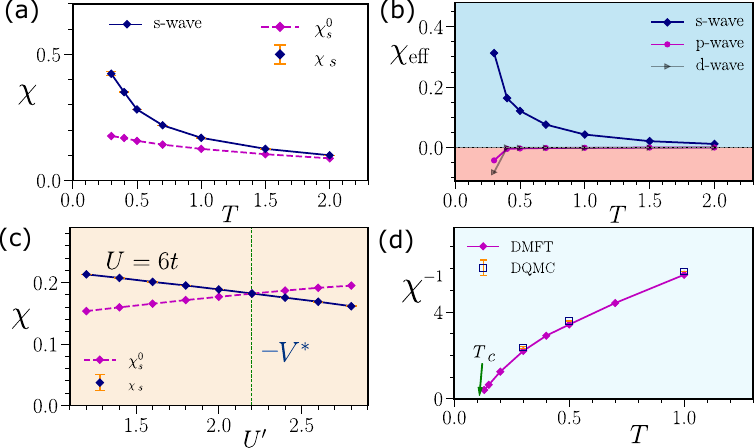}
\caption{Pairing susceptibilities $\chi$ and effective pairing attractions $V^{*}$ of the spinless trilayer Hubbard model.\textbf{(a): }Pairing susceptibilities
$\chi_{l} $ for $\ell=s-$ wave  as a function of temperature $T$
at $U=6,U^{*}=0,n=0.5$. The bubble contributuion $\chi^{0}_{s}$ is shown in dashed line. \textbf{(b):
}Effective pairing susceptibilities $\chi_{\mathrm{eff}} = \chi_{l}  - \chi_{l}^{0} $ as a function
of  $T$. For $p-$ and $d-$ wave, $\chi_{\mathrm{eff}} <0 $ , denoting both pairing symmetries are unfavoured. Other parameters are the same as in (a). \textbf{(c):
} $\chi_s$ and $\chi_s^{0}$ as functions of $U^{\prime}$  
at $U=6t,n=0.5, T=0.5t$. The critical $U^{\prime}_c$ where $ \chi_{eff} \equiv \chi_s - \chi_s^{0} \approx 0$ is used to define the effective pairing attraction $V^{*} = -U^{\prime}_c $. 
\textbf{(d): } Inverse $s-$ wave pairing susceptibility as a function of temperature $T$. Here $U=6t, n=0.6$. $\chi^{-1} \rightarrow 0$ is indicated as where $T$ approaches superconducting transition temperature $T_c$. Here $U^* = 0$ and $n_1 = n_2=n_3=n$ for all plots.
Results are obtained by $6\times 6\times 3$ DQMC simulation.
} 
\label{fig:chi}
\end{figure}

\textit{weak-  and strong-coupling limits - } To gain insight from an analytic perspective, we begin by considering the spinless model. Without loss of generality, we set $U_{12}=U_{23}=U$, and $U_{13}=U^{*}=0, U(r)=0$, \textit{i.e.,} neglecting long-ranged repulsions as well as the direct repulsion between the top and bottom layers. In  weak-coupling limit $U\ll  t$, second order KL Feynman diagrams yield an inter-layer attraction $V_{\mathrm{eff}}$ between the top- and bottom- layers within the same unit cell, mediated by the polarization of the middle-layer ($l=2$), as illustrated in Fig.~\ref{fig:illu}b. In the instantaneous limit, $V_{\mathrm{eff}}$ scales like (see End Matter),
\begin{eqnarray}
V_{\mathrm{eff}} & \propto & -U^{2}n_{2} (1-n_{2})/t
\label{eq:vweak}
\end{eqnarray} 
where  $n_{2}$ is the electron density of the middle-layer.
In the large $U/t >>1 $ limit at half-filling, to avoid repulsion between electrons (holes) on neighbouring layers, if the top- and bottom- layer are occupied by a pair of electrons (holes),
the middle-layer will be occupied by a hole (electron). This occupation configuration has an energy gain being of the order of $U$, hence generating an effective attraction $V_{\mathrm{eff}}$  between top- and bottom- layer electrons,
\begin{eqnarray}
 V_{\mathrm{eff}} \propto -U. 
 \label{eq:vstrong}
\end{eqnarray} 

see  Fig.~\ref{fig:illu}c and End Matter. Therefore, in both weak-coupling and strong-coupling regimes,  an effective attraction between top- and bottom- layer electrons can be generated from the many-body effects of inter-layer repulsion $U$, which eventually leads to superconductivity in the system.


\textit{DQMC result -}
We  now use numerical exact DQMC to study the pairing attraction. We first focus on spinless model with $U_{12}=U_{23}=U,U_{13}=U^{*}=0, U(r)=0$. 
In Fig.~\ref{fig:chi}a, we  present the pairing susceptibility $\chi = \chi_{\ell}^{(l,m)}$ for inter-layer ($l=1, m=3$)
 $s$-wave pairing (see End Matter) at $U=6, n=0.5$. One can see that as temperature $T$ decreases, $\chi_{s}^{(1,3)}$ (solid line) grows rapidly. 
The notorious sign problem  prevents DQMC from accessing the Berezinskii-Kosterlitz-Thouless (BKT) temperature $T_{BKT}$  for a diverging $\chi$.
A pragmatic approach to tackle the pairing tendency in DQMC is to estimate the effective pairing $\chi_{\mathrm{eff}}=\chi-\chi_{0}$,
where $\chi$ and $\chi_{0}$  are respectively the full  pairing susceptibility and bubble contributions.
The sign of  $\chi_{\mathrm{eff}}$ reflects whether the vertex corrections, \textit{i.e.}, the interaction
effects beyond single-particle level, enhance superconductivity ($\chi_{\mathrm{eff}}>0$
) or suppress superconductivity ($\chi_{\mathrm{eff}}<0$ ). Fig.~\ref{fig:chi}a and Fig.~\ref{fig:chi}b
 suggests that the $s-$wave pairing
are  favoured by correlation effects , whereas 
the $p-$ and $d-$ wave are unfavoured.  We have
also considered other pairing correlations, such as the $s-$,$p-,d-$ wave pairings between
the top- and middle- layers [$(l,m)=(1, 2)$], where they are all unfavoured (data not shown).

To quantify the attraction between pairing electrons, we compute a quantity $V^{*}$ in DQMC, which is defined by probing the system's  $\chi_{\mathrm{eff}}$ response to pair-breaking repulsion.
Specifically, we calculate  $\chi_{\mathrm{eff}}$ for $s-$ wave pairing between top- and bottom- layer
electrons while introducing an explicit interlayer repulsion $ U^{'}$ between them. For $U^{'}=0$ , one finds $\chi_{\mathrm{eff}}>0$,
indicative of favoured pairing, like shown in Fig.\ref{fig:chi}b. Increasing  $U^{'}$ will suppress $\chi_{\mathrm{eff}}$, and  a critical  $U_{c}^{'}$ 
can be identified when  $\chi_{\mathrm{eff}}\approx0$. We then define the effective pairing attraction
 $V^{*}$ as $V^{*}=-U_{c}^{'}$, as shown in Fig.~\ref{fig:chi}c. This quantity naturally measures the binding strength between two pairing electrons. As shown in Fig.~\ref{fig:illu}d for the spinless model,  our DQMC result of $V^{*}$ exhibits  
a  quadratic $U$ dependence, $V^* \propto U^2/t$ for small $U$ ($U \lesssim 5t$), in nice agreement with Eq.~\ref{eq:vweak} that complies with the conventional KL theory~\cite{kohn1965new,romer2020pairing}
(see also End Matter).

 For larger $U \gtrsim 5t$,  the  quadratic KL  relation  
 no longer strictly holds, $V^{*}$ gradually  transits to a new linear scaling, $V^{*}\propto -U$ (dashed line in Fig.~\ref{fig:illu}d), underscoring the fundamental difference between
 KL pairings at strong-couling and that at weak-coupling.
Above results indicate that in both large and small $U$ limits, DQMC result of $V^{*}$ agrees well with our analytic analysis on $V_{\mathrm{eff}}$
in Eq.~\ref{eq:vweak} and Eq.~\ref{eq:vstrong}.
In essence, 
the overall  growing tendency of $V^{*}$ with $U$ suggests that one can keep increasing
inter-layer Hubbard repulsion $U$ to realize a larger $V^{*}$ between electrons,  hence potentially a higher $T_{c}$ of SC.

\begin{figure}
\includegraphics[scale=0.42]{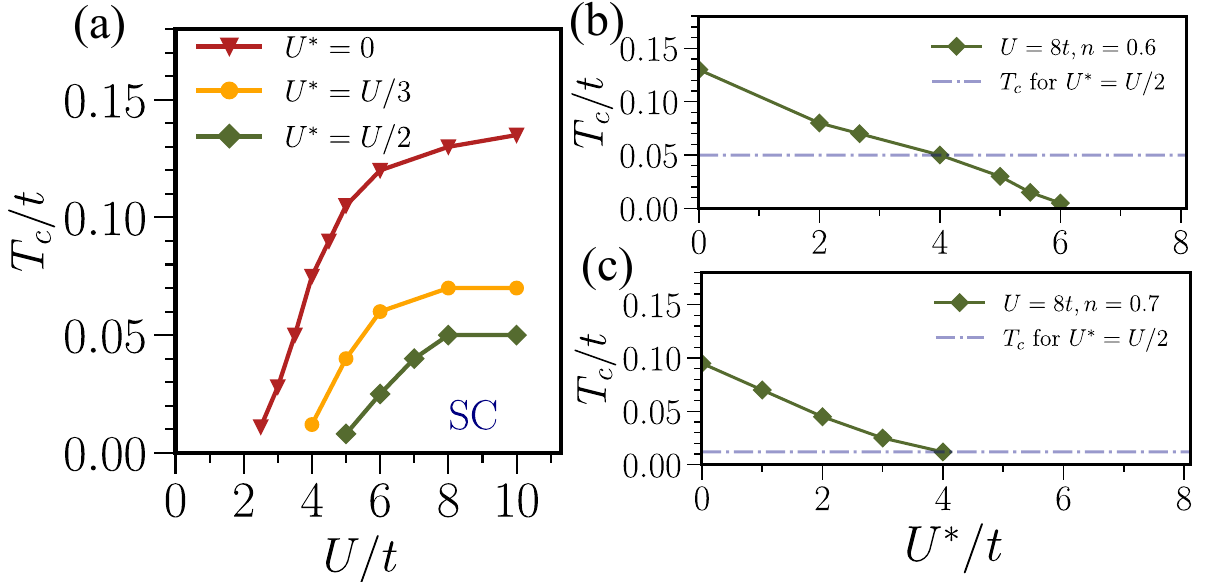}
\caption{Superconducting transition temperature $T_c$ from DMFT calculations. \textbf{(a)} $T_c$  as a function of $U$  at $ n_1=n_2=n_3= 0.6$ for  a few different $U^*$. \textbf{(b)} $T_c$ as a function of $U^*$ at $ U = 8t$ at $n=0.6$  (20\% electron doping). \textbf{(C)} Same as (b) but at $n=0.7$ (40\% electron doping). In (b) and (c) The values of $T_c$ at $U^* = U/2 = 4t$ is marked out by dashed lines.
}
\label{fig:tc}
\end{figure}

\textit{SC instability-}
We now use CDMFT to compute the superconducting $T_{c}$  at low temperatures. $T_c$
 is identified as the temperature where pairing susceptibility $\chi$ diverges in CDMFT.
 In Fig.~\ref{fig:chi}d, we plot the inverse of  $s-$ wave pairing susceptibility,  $\chi ^{-1}$ as a function of $T$ at $U=6, n=0.6$.  DQMC data available at high temperatures are also presented, where excellent agreement between DMFT and DQMC results can be seen. Extrapolating CDMFT result of $\chi ^{-1}$ to zero, $\chi ^{-1} \rightarrow 0$, we obtain a superconducting $T_c \approx 0.12t$ for the given parameters in this plot. Repeating the calculations of $T_c$
 for different $U$, the $U$-dependence of $T_c$  is obtained in Fig.~\ref{fig:tc}a. In this plot, one sees that $T_c$ (triangles) grows with $U$ in a broad parameter
 range,  in consistent with the growing tendency of $V^*$ in DQMC result ( Fig.~\ref{fig:illu}d). Specifically, escalated $T_c \gtrsim 0.12t$ can be found when $U$ is in the range of  $ U \sim (6t,10t)$, namely, when $U$ becomes comparable with bandwidth $W$ ($W=9t$ for 2D triangular lattice), KL SC can be greatly enhanced. For comparison, in the magnetic-fluctuation mediated SC in the single-band 2D Hubbard model, the optimal superconducting $T_c$ obtained by CDMFT is commonly less than $T_c \lesssim 0.05t$\cite{fratino2016organizing,maier2019pairfield,liu2025interplay}. This means that in terms of the reduced pairing energy scale, $T_c/t$, the strong-coupling KL SC in the trilayer Hubbard model has an intrinsically stronger pairing force than that in the magnetic-fluctuation mediated SC in Hubbard model.

 \begin{figure}[b]
\centering{}\includegraphics[scale=0.32]{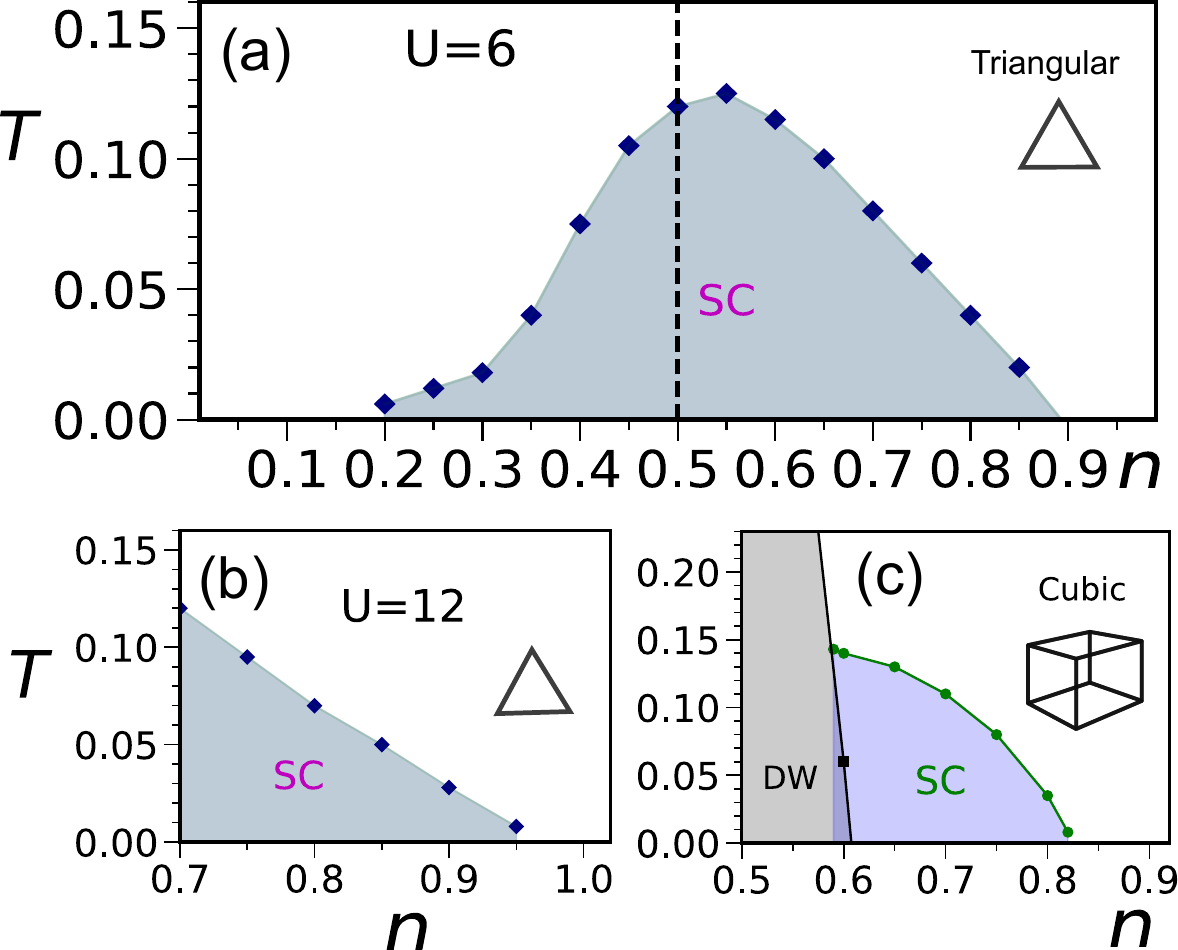}\caption{Phase diagrams showing 
superconducting instability (SC) and density wave (DW) instability.
\textbf{(a)} Superconducting transition temperature $T_c$ as a function of electron density $n$ on the 2D triangular lattice at $U=6t$.
(b) Same as (a) but at $U=12t$.
\textbf{(c)} Density wave and superconducting transition temperatures as a function of electron density $n$ on the 
3D cubic lattice at $U=10t$. 
Here for all three cases $n_{1}=n_{2}=n_{3}=n$. The repulsion between layer- (pseudo-spin ) 1 and layer- (pseudo-spin) 3,  $U^{*}=0.$}
\label{fig:pd}
\end{figure}

 Furthermore,  the nonzero $U^{*}$ cases  shown in Fig.~\ref{fig:tc}a (dots and diamonds) suggest that in the presence of  significant remnant Coulomb repulsion ($U^{*} = U/3, U/2$ ) between paired electrons,  large superconducting $T_c$ can be still maintained. In  Fig.~\ref{fig:tc}b and  Fig.~\ref{fig:tc}c, $T_c$  as a function of $U^{*}$ are shown at $n=0.6,0.7$ respectively (for fixed $U=8t$),  where finite $T_c$ can be found even  when $U^{*} \geq U/2$. Specifically, for the $n=0.6$ case, an enhanced $T_c \approx 0.05t$ is found at $U=8t, U^* =4t$. This means that 
 in the large screening length limit $\lambda_{TF} \rightarrow \infty$ along $c$-axis, \textit{i.e.}, when Coulomb potential in the vertical direction becomes $U(r_z) \propto \frac{e^{-r_z/\lambda_{TF}}}{r_z} \sim \frac{1}{r_z}$, (thus $U^* \sim U/2$ if layer distances $d_{12} = d_{23}$),
 the strong-coupling KL SC can still survive. This finding lie beyond  the expectation of high-temperature DQMC results presented in  Fig.~\ref{fig:illu}d, where $V^{*} \sim -U/2$ for large $U$. In a purely static picture,
 $U^{*} \gtrsim U/2$ will lead to a net repulsion between top- and bottom-layer electrons [ \textit{i.e.}, $ (|V^{*}| - U^{*} ) \lesssim 0 $],  thereby severely inhibiting pairing between them. We attribute the persistence of high $T_c$ under  $U^{*} \gtrsim U/2$ condition to the retarded pairing or suppression of layer density fluctuations (see discussions in End Matter).
A further remarkable point to emphasize is the sizeable  $T_c$ up to a filling factor of $n=0.7$ (40\% electron doping), as shown in Fig.~\ref{fig:tc}c. This presents a clear difference with  cases in  Hubbard model for cuprates~\cite{fratino2016organizing}, where SC rapidly vanishes beyond 
 $p \lesssim 30\%$  for hole-doping and 
$p \lesssim   20\%$  for electron-doping. This distinctness is rooted in the ``pairing glue": SC in cuprates relies on the collective spin fluctuations~\cite{dong2022quantifying}, which are suppressed at large dopings. In contrast, the strong-coupling KL SC identified here does not involve any intermediate collective modes, thus ensuring robustness across a broad  range of $n$.

\textit{The phase diagrams -}
 Fig.~\ref{fig:pd}a presents a superconducting phase diagram showing 
$T_{c}$ as a function of electron density $n$
for the spinless model on 2D triangular lattice from nearly empty ($n\sim 0.2$) to nearly full-filled ($n\sim 0.9$) at
$U=6t, U^{*} = 0$. One sees that 
 $T_c$ is maximumized as $n$ being close to half-filling. Again, here  we find SC instability in an extremely large range of $n$. Increasing  repulsion $U$ can even further expand the SC regime. For instance, on the electron doped side, finite $T_c$ can be found for $n_c \gtrsim 0.95$ at $U=12t$ ( Fig.~\ref{fig:pd}b).
 
\begin{figure}[t]
\centering{}\includegraphics[scale=0.4]{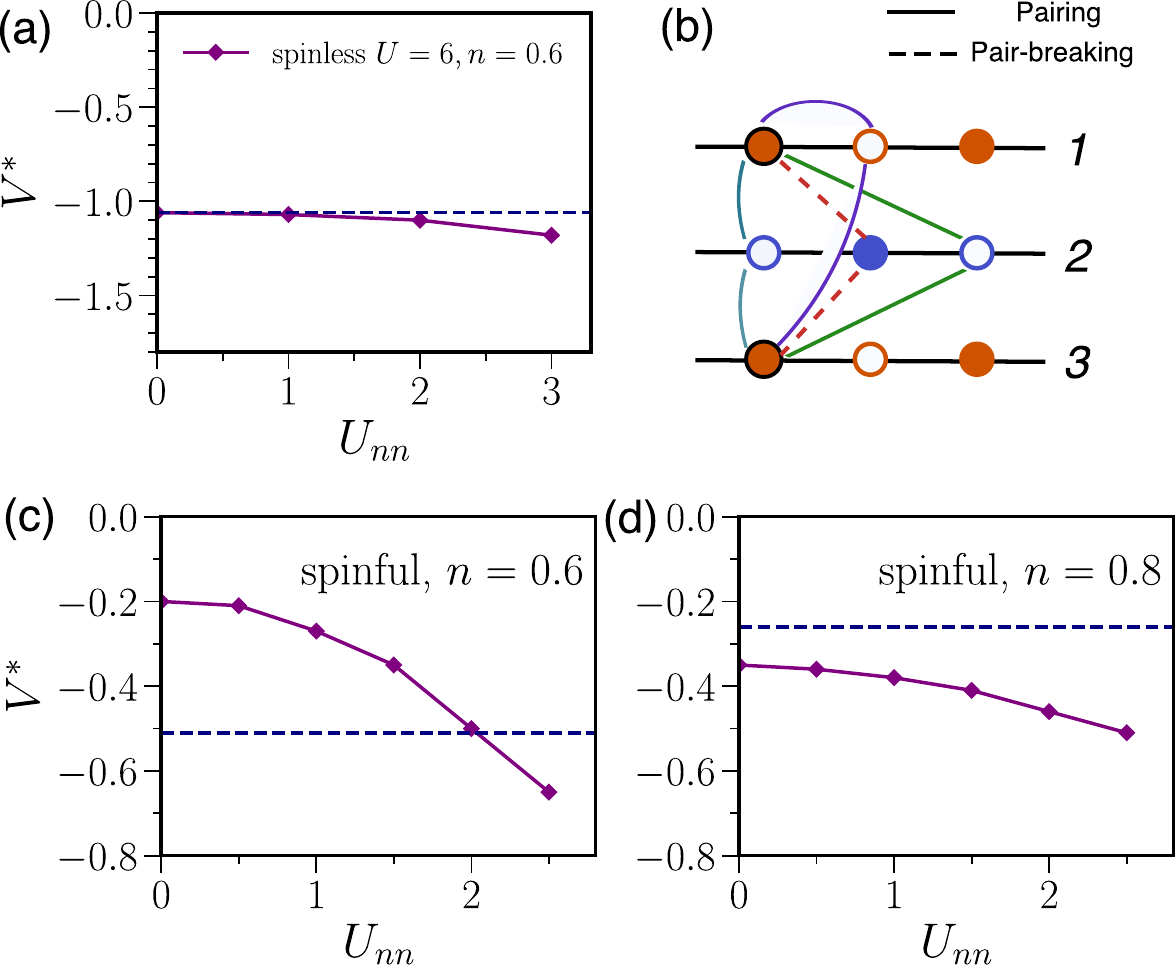}
\caption{Long-range Coulomb repulsions can enhance pairing attraction $V^*$. $U_{nn} = U_{ll}(\vec{r_1}) $ denotes the intra-layer NN repulsion, $|| \vec{r_1} ||_{xy} =1$.\textbf{(a)} $V^*$ as a function of  $U_{nn}$ for  spinless model at $T=1.0$. Other NN repulsions are set as $U_{12} (\vec{r_1}) = U_{23} (\vec{r_1}) = U_{nn}/2 $, $U_{13} (\vec{r_1}) = U_{nn}/4$.  \textbf{(b)} A staggered distribution of electrons at three layers minimizing the potential energy is shown in 1D chain, on which 
interaction terms that promote (solid lines)  or suppress (dashed lines) inter-layer pairing is illustrated. Filled (hollow) symbols denote electrons (holes).
 \textbf{(c)} and  \textbf{(d)}:  $V^*$ as a function of  $U_{nn}$ for the  spinful model at $U_o = 8, U=4,  U^{*} =0, T=1.0$. Inter-layer NN repulsion  $U_{lm}(\vec{r_1}), l\neq m $ is neglected in (c) and (d)  to avoid sign problem.
}
\label{fig:ur}
\end{figure}

In vicinity of half-filling, we find strong density fluctuation in CDMFT calculations which tends to form staggered distribution of electrons at the three layers ( see Fig.~\ref{fig:ur}b ).
However, due to the geometry frustration, no true long-range ordered state is found on triangular lattice. 
 For comparison, in Fig.~\ref{fig:pd}c, we present a similar phase diagram for the non-frustrated cubic lattice (where the layer index $l$ now can be understood as a pseudo-spin index). At half-filling ($n=0.5$), this system indeed develops a density
wave (DW) phase.
This long-range  DW  instability competes with SC order, which can be destroyed by doping electrons (holes). As DW order vanishes (e.g. $n \gtrsim 0.6$ in Fig.~\ref{fig:pd}c), SC instability emerges. Further increasing $n$ leads to the decreasing of superconducting $T_c$, similar to the triangular lattice result.
This finding demonstrates that strong-coupling KL SC is robust against changes in lattice geometry (bipartite V.S. non-bipartite) and dimensionality (3D V.S. 2D).


 \textit{Fffects of $U(r)$ and spinful model-}
 Considering long-ranged Coulomb repulsions beyond a unit cell, there can be different $U_{lm}(r)$ terms  stabilizing or destabilizing the binding of electron (hole) pairs at top- and bottom- layers, as illustrated in Fig.~\ref{fig:ur}b. We find that the overall effects of the nearest-neighboring (NN)  repulsion enhances $V^{*}$, as shown in  Fig.~\ref{fig:ur}a for the spinless model. 
 
For the spinful model of Eq.~\ref{eq:hamil}, twofold competing effects can influence interlayer KL pairing relative to the spinless case: (1) the extra spin degree of freedom in the middle layer can enhance pairing by offering an additional polarization channel; (2) on the other hand, strong on-site repulsion $U_o$ between  opposite-spin electrons induces  local electron–hole pairs ($i.e.$, local moments), which is detrimental to interlayer electron-electron pairing.
The balance between these two opposing mechanisms then affects $T_c$ in the spinful model.

We examine  influence of both  on-site $U_o$ and NN repulsion $U_{nn}$ on $V^*$ in Fig.~\ref{fig:ur}c and Fig.~\ref{fig:ur}d at $U_o = 8t, U=4t$.
Fig.~\ref{fig:ur}c shows that at $U_{nn}=0$, 
the spinfull model yields a smaller $V^{*}$ compared to the corresponding spinless model ( dashed line, without $U_o$, and $U=4t, U_{nn} = 0$), confirming that $U_o$  suppresses KL pairing.  As $U_{nn}$ increases, $V^{*}$ grows and exceeds the spinless value once $U_{nn} \gtrsim 2t$. We note that long-range Coulomb repulsions can promote charge density wave (CDW) fluctuations, hence suppressing the detrimental local moments induced by $U_o$. This could be another route that $U(r)$ enhances $V^*$ in spinful cases.
At a higher filling $n=0.8$ (Fig.~\ref{fig:ur}b), $V^{*}$ is also seen enhanced by $U_{nn}$. In this case, the spinful model already gives a larger $V^{*}$ than its spinless counterpart even at $U_{nn} =0$. 
This is expected, because, far away from half-filling, local moments are suppressed by doping, thus the additional polarization channel effect in spinful model becomes predominant, which results in a larger $V^*$.

\begin{figure}[t]
\centering{}\includegraphics[scale=0.34]{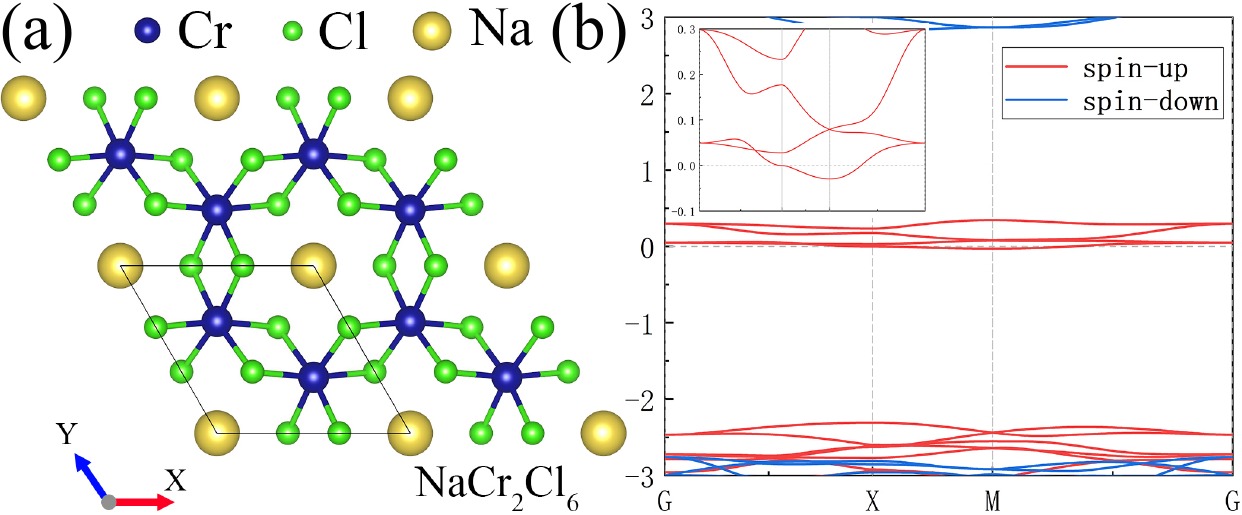}
\caption{Crystal and band structure of $\mathrm{Na}$-adsorbed transition metal halides   
$\mathrm{CrCl_3}$. \textbf{(a)} Crystal structure of $\mathrm{Na}$-adsorbed $\mathrm{NaCr_2Cl_6}$.
\textbf{(b)}  Energy band structure of $\mathrm{NaCr_2Cl_6}$. \textbf{Inset}: Magnified view of energy bands near Fermi level.
}
\label{fig:matt}
\end{figure}

 \textit{Realization of strong-coupling KL pairing-}
 As demonstrated above, enhanced KL pairing can be achieved in a trilayer structure of 2D electrons when  $U \sim W$. 
In 2D van der Waals layered materials, the interlayer repulsion can reach  $U \lesssim 0.5 eV$ (see End Matter and Ref.~\cite{interlayU}). To satisfy the strong-coupling condition, a narrow electronic bandwidth,  $W \lesssim 0.5eV$,  is therefore needed.  Through density functional theory (DFT) and constrained random-phase approximation (cRPA) calculations, we have identified several atomically thin two-dimensional half-metals that can effectively realize the trilayer spinless Hubbard model fulfilling $W \lesssim U$. Examples include sodium-adsorbed transition-metal trihalides such as $\mathrm{NaCr_2{C}l_{6}}$, where  $W \sim 0.087eV$ and $U \sim 0.3 eV$ (see Fig.~\ref{fig:matt} and End Matter), and  an iron-doped phosphorene, $\mathrm{FeP_{35}}$~\cite{liu2015black} (see End Matter). We note that the transition metal dichalcogenide(TMD)-based  moir\'e superlattices~\cite{zhai2025twistronics} exhibit bandwidths below  $10 \mathrm{meV}$ (inter-site hopping strength $t \sim 1 \mathrm{meV}$)~\cite{wufc2018}, which is smaller than the inter-layer repulsion~\cite{xie2022strong}. Hence these systems may also be suitable for implementing our proposal.  Finally, recent advances in engineering artificial 2D electron lattices within semiconductor heterostructures have  yielded bandwidths of only a few meV~\cite{wang2025artificial}, offering another promising platform for realizing strong KL superconductivity.

\textit{Discussion and conclusion - }
We have proposed a scheme for realizing a novel emergent quantum many-body state: strong-coupling KL superconductivity in 2D electrons. Recently, novel exciton-mediated SC in atomically thin
VdW heterostructures have been proposed~\cite{von2024sc,zerba2024realizing}.
 A key distinction of our proposal lies in the large pairing energy scale $V^*$ that does not require any collective mode to mediate the pairing. Consequently, it exhibits robustness against variations in lattice geometry and dimensionality, and can endure a large remnant Coulomb repulsion  $U^{*}$ between pairing electrons. These advantages allow substantial flexibility to control SC by tuning sample thickness, dielectric environment or interlayer distance in VdW materials.
Although our low-temperature DMFT calculations of $T_c$ are restricted to the spinless model for  half-metals due to numerical challenges, the DQMC results do suggest that the spinful model (for normal 2D metals) can, under proper conditions,  host an even stronger effective pairing attraction  $V^*$.

In summary, we have demonstrated the emergence of a robust, strong-coupling inter-layer $s$-wave Kohn-Luttinger superconducting state in multi-layer Hubbard models prototyping stacked 2D electron systems. Our \textit{ab initio} calculations identify a few candidate materials capable of realizing this novel state, indicating that 2D van der Waals materials can also constitute a versatile platform for designing and manipulating unconventional superconductivity.

\textit{Acknowledgment} We are grateful to Wenchen Luo and Fadi Sun for useful discussions. This work is supported by the National Natural Science Foundation of China (Grants No.12274472, No. 12494594). We also thank the support from the Research Center for Magnetoelectric Physics of Guangdong Province (Grants No. 2024B0303390001) and Guangdong Provincial Quantum Science Strategic Initiative (Grant No. GDZX2401010). H.Y. acknowledges support by Guangdong Provincial Quantum Science Strategic Initiative (Grant No. GDZX2501003) and NSFC under Grant No. 12274477.

\bibliography{kl.bib}

\onecolumngrid

\section{END MATTER}

\twocolumngrid

\subsection{Multi-layer Hubbard Model and Methods}

In the trilayer spinless model, we consider the case where  $U_{12}=U_{23}=U$ and $U_{13}=U^{*}=0$, corresponding to complete screening of the Coulomb repulsion between the top- and bottom- layers. In the weak-coupling limit $U\ll  t$, the second order KL Feynman diagram yield an pairing attraction between the top- and bottom- layer electrons, mediated by polarization processes in the middle-layer ($l=2$) electrons~\cite{kohn1965new,romer2020pairing},

\begin{eqnarray}
V_{KL}(\bm{k}, \bm{k'}) = \frac{U^2}{N}[\chi_0( \bm{k}+ \bm{k'})-\chi_0( \bm{k}-\bm{k'})] \\
\chi_0(\bm{q}) = \frac{1}{N}\sum_{\bm{k}}g_0( \bm{k}+\bm{q})g_0( \bm{k})
\end{eqnarray}
where $\chi_0( \bm{q})$ is the Lindhard function at transfer momentum $\bm{q}$, and $N$ is the number of unit cells. In the instantaneous limit, this Kohn-Luttinger interaction introduces an effective attraction $V_{\mathrm{eff}}$ between a pair of electrons located
on top- and bottom- layer  in a unit cell  in real-space that scales like,
\begin{eqnarray}
V_{\mathrm{eff}} & \propto & -U^{2}n_{2} (1-n_{2})/t
\label{eq:vweak2}
\end{eqnarray} 
where $n_{2}$ is the electron filling factor of the middle-layer.

\begin{figure}[b]
\centering{}\includegraphics[scale=0.34]{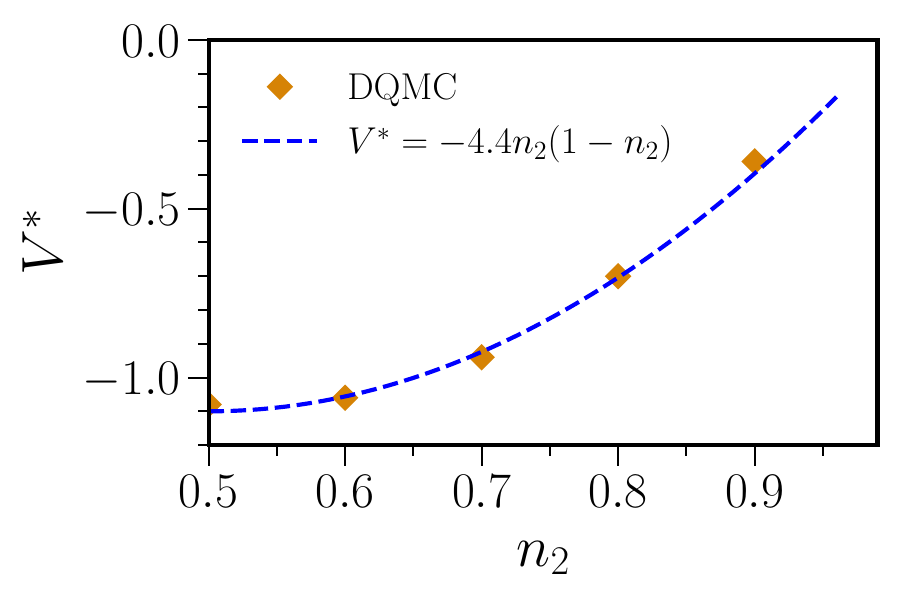}
\caption{ Effective pairing attraction $V^*$ as a function of  the density of the middle-layer $n_2$ in the trilayer spinless model. Here $n_1 = n_3 = 0.5, U=4t, U^* =0, U(r) =0, T=0.5t$ are fixed.  $6\times 6\times 3$ lattice is used in DQMC simulation.
}
\label{fig:vn}
\end{figure}

In Fig.~\ref{fig:illu}d in the main text, we demonstrated that the DQMC result indeed follows a  $V^* \sim U^2/t$ scaling for $V^*$ at small $U$. Additionally, Fig.~\ref{fig:vn} illustrates $V^*$ as a function of $n_2$, where the relation $ V^* \propto n_2(1-n_2)$  is also clearly satisfied.

In the large $U$ limit ( $U/t >>1 $ or  $U/W >>1 $) at half-filling, to avoid repulsion $U$ between electrons (holes) on neighboring layers, the middle-layer tends to be occupied by the opposite carrier of the top- and bottom-layer (see also Fig.~\ref{fig:ur}b). This occupation configuration will have an energy gain of the order of $-U$ over the first excited state. To be specific, we consider three layers of intra-layer electron-hole pairs (denoted as $\bb-\cc$) in the spinless model. There are three different states of most interest to our problem,
\begin{widetext}
\begin{eqnarray*}
\frac{\sqrt{2}}{2} \left( \left\vert
\begin{array}{c}
\bb-\cc \\ \cc-\bb \\ \bb-\cc
\end{array}
\right\rangle
+ \left\vert
\begin{array}{c}
\cc-\bb \\ \bb-\cc \\ \cc-\bb
\end{array}
\right\rangle \right) \quad \textcircled{A} \\
\frac{1}{2} \left( \left\vert
\begin{array}{c}
\cc-\bb \\ \cc-\bb \\ \bb-\cc
\end{array}
\right\rangle
+ \left\vert
\begin{array}{c}
\cc-\bb \\ \bb-\cc \\ \bb-\cc
\end{array}
\right\rangle
+ \left\vert
\begin{array}{c}
\bb-\cc \\ \cc-\bb \\ \cc-\bb
\end{array}
\right\rangle
+ \left\vert
\begin{array}{c}
\bb-\cc \\ \bb-\cc \\ \cc-\bb
\end{array}
\right\rangle \right) \quad \textcircled{B} \\
\frac{\sqrt{8}}{8} \left( \left\vert
\begin{array}{c}
\cc-\bb \\ \bb-\bb \\ \cc-\cc
\end{array}
\right\rangle
+ \left\vert
\begin{array}{c}
\bb-\cc \\ \bb-\bb \\ \cc-\cc
\end{array}
\right\rangle
+ \left\vert
\begin{array}{c}
\cc-\bb \\ \cc-\cc \\ \bb-\bb
\end{array}
\right\rangle
+ \left\vert
\begin{array}{c}
\bb-\cc \\ \cc-\cc \\ \bb-\bb
\end{array}
\right\rangle
+ \textcircled{1} \Longleftrightarrow \textcircled{3} \right) \quad \textcircled{C}
\end{eqnarray*}
\end{widetext}

 In state $\textcircled{A}$, electrons can avoid any inter-layer repulsion $U$ between neighboring layers, thus it becomes the ground state in the large $U$ limit at half-filling. Obviously,
 this state is the physical origin where pairing of two electrons emerges on top- and bottom-layer respectively.  State $\textcircled{B}$ is the first excited state, given the total density fluctuation in each layer is forbidden (\textit{i.e.} $\delta n_{l} = 0$). This state has an energy gain of $\Delta E = (U-t)$ relative to the ground state  $\textcircled{A}$. On the other hand, if the total density fluctuation in layers are allowed, then state  $\textcircled{C}$ becomes the first excited state, which has an energy gain of $\Delta E = (U/2-t)$ over the ground state  $\textcircled{A}$ [$\textcircled{1} \Longleftrightarrow \textcircled{3}$ means interchange the top- and bottom- layer configurations of the first four kets]. 
 As a result, at half-filling 
in the strong-coupling limit $U/t \rightarrow \infty$, the pairing attraction between top- and bottom-layer electrons $V_{\mathrm{eff}}$ in the spinless model should in general scale like,
\begin{eqnarray}
 V_{\mathrm{eff}} \propto -U. 
 \label{eq:vstrong2}
\end{eqnarray} 
In DQMC calculations at high temperatures, layer density fluctuations can prevail. Thus it is reasonable finding out a scaling of ,
\begin{equation}
V^* \sim -U/2
\end{equation}
in Fig.~\ref{fig:chi}d At low temperatures, [or if the intra-layer nearest-neighbor (NN) repulsion $U_{nn}$ is introduced], the layer density fluctuations $\delta n_{l}$ can be suppressed, which may lead to a larger $\mathrm{V_{eff}}$
\begin{equation}
\mathrm{V_{eff}} \sim -U
\end{equation}
Our low-temperature DMFT calculations show that robust KL pairing can persist  when  $U^* \gg U/2$ (Fig.~\ref{fig:tc}b), which implies an effective interaction  $|V_{\mathrm{eff}}| \gg U/2$. This phenomenon likely stems from the suppression of density fluctuations, as discussed above. However, the contribution from retardation effects of pairing 
at large $U^*$ cannot be ruled out.
Further investigation is required to  fully clarify this point.

In DQMC simulation, the pairing susceptibility $\chi_{\ell}^{(\alpha,\beta)}$ is defined as,

\begin{eqnarray*}
\chi_{\ell}^{(\alpha,\beta)}=\frac{1}{N_{s}N_{\ell}}\sum_{i,j,a,b}\int_{0}^{\beta}d\tau\langle\mathcal{T}_{\tau} p_{j,a}^{\dagger \ell,(\alpha,\beta)}(\tau)p_{i,b}^{\ell,(\alpha,\beta)}(0)\rangle
\end{eqnarray*}

here $p_{i,b}^{\ell,(\alpha,\beta)}$ is the pairing operator for a pair of electrons located
at sites $i$ and $i+b$. For brevity, here we use composite index $\alpha \equiv (m,\sigma)$ to label electron components.  For a spin (layer) singlet/triplet pairing,
$p_{i,b}^{\ell,(\alpha,\beta)}=\frac{\sqrt{2}}{2}f_{i,b}^{\ell}(c_{i,\alpha}c_{i+b,\beta} \pm c_{i,\beta}c_{i+b,\alpha})$,
where $f_{i,b}^{\ell}$ is the form factor in real space associated
to specific pairing symmetry $\ell$. $N_{s}$ denotes the number of sites,
and $N_{\ell}= \sum_{b}|f_{i,b}^{\ell}|^2 $ is a renormalization factor.

Our DMFT calculations are based on a three-site effective impurity cluster representing a single unit cell (top-, middle-, and bottom- layer).
 The impurity model was solved using an open-source implementation of the continuous-time quantum Monte Carlo (CTQMC) method within the TRIQS 3.2 framework~\cite{CTHYB2016,TRIQS2015}, and the Hirsch-Fye quantum Monte Carlo impurity solver.

\begin{table*}
\caption{Bare $V$, Hubbard $U_o$, repulsion between electrons at neighboring layers $U_{12}$, orbitals, Energy window $E_{w}$ for Wannier projection, and number of layers $N_{m}$. Here, the Fermi energy $E_{F}$ is set to 0 and all the energy units are in eV. $\mathrm{FeP_{7/15}}$ are Fe doped in $\mathrm{1 \times 2 \times 1}$ and $\mathrm{2 \times 2 \times 1}$ phosphorene supercell respectively. For projected orbitals, $\mathrm{P_{01}}$ refers to the P atom adjacent to the doped atom. }
\setlength{\tabcolsep}{15pt}
  \label{tab:booktabs}
    \centering
    \begin{tabular}{ccccccc} 
        \toprule 
        Materials & $V$ & $U_o$ & $U_{12} \equiv U$ & orbitals & $E_{w}$ & $N_{m}$ \\
        \midrule 
\\
       $\mathrm{Cr_{2}I_{6}}$ & 14.643 & 2.89 & /  & Cr($d_{z^{2}},d_{xy},d_{x^{2}-y^{2}}$) & [-0.48,0.52] & 1 \\
             & 15.04 & 2.86 & 0.33 &  & [-0.48,0.52] & 2 \\
\\
      $\mathrm{Cr_{2}Br_{6}}$ & 15.65 & 3.53 & / & Cr($d_{z^{2}},d_{xy},d_{x^{2}-y^{2}}$) & [-0.643,0.857] & 1 \\
             & 16.12 & 3.72 & 0.45 &   & [-0.643,0.857] & 2 \\
\\
      $\mathrm{Cr_{2}Cl_{6}}$ & 16.36 & 4.23 & / & Cr($d_{z^{2}},d_{xy},d_{x^{2}-y^{2}}$) & [-0.46,0.54] & 1 \\
             & 16.69 & 4.5 & 0.57 &   & [-0.46,0.54] & 2 \\
\\
       $\mathrm{FeP_{7}}$ & 3.38 & 0.742 & / & P$_{01}$($p_{z},p_{x}$) & [-1.85,0.65] & 1 \\
             & 4.1 & 0.684 & 0.152 &   & [-1.91,1.09] & 2 \\
        $\mathrm{FeP_{15}}$ & 3.84 & 0.524 &/ & P$_{01}$($p_{z},p_{x}$) & [-0.967,1.033] & 1 \\

        \bottomrule 
    \end{tabular}
\end{table*}

\subsection{\textit{ab initio} Calculations}

\begin{figure}[b]
\centering{}\includegraphics[scale=0.34]{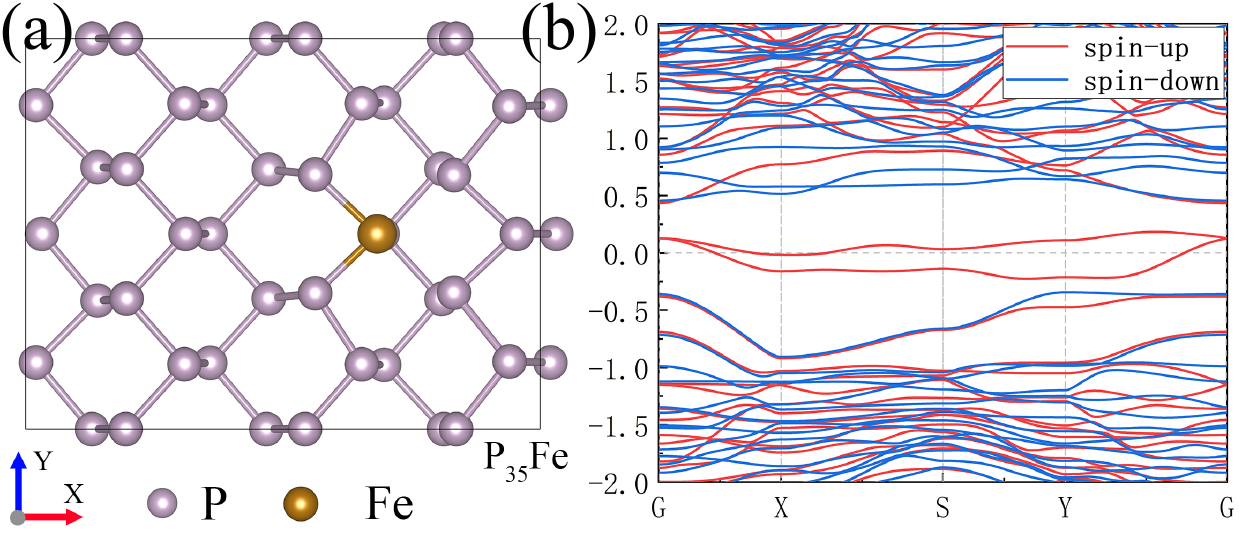}
\caption{Crystal and band structure of $\mathrm{Fe}$-adoped phosphorene.  \textbf{(a)} Crystal structure of $\mathrm{Fe}$-adoped $\mathrm{FeP_{35}}$.
\textbf{(b)}  Energy band structure of $\mathrm{FeP_{35}}$.
}
\label{fig:PFe}
\end{figure}

In the first principles density functional theory (DFT) calculations of  $\mathrm{NaCr_2{C}l_{6}}$ and iron-doped phosphorene, $\mathrm{FeP_{35}}$ (Fig.~\ref{fig:PFe}), we employ the Vienna Ab initio Simulation Package (VASP)~\cite{kresse1993ab,kresse1996efficient} and selected the Perdew-Burke-Ernzerhof (PBE) exchange-correlation functional~\cite{perdew1996generalized} for structural relaxation and band structure calculations. To ensure a balance between calculation accuracy and efficiency, the plane-wave cutoff energy was set to 600 eV. The convergence criteria for electronic optimization and structural relaxation were strictly set to $10^{-6}$ eV and 1 meV/Å respectively.
To describe better the transition metal atoms, we use DFT+$U$ method with U = 3 eV for Cr and U = 3.5 eV for Fe~\cite{shi2022structural,naveas2023first} in calculating the band structure. 

To estimate the Hubbard interaction terms, in particular the inter-layer repulsion $U$ between neighbouring layers, we have employed bilayer systems ($N_m = 2$) in the constrained random phase approximation (cRPA) calculation~\cite{gao2025topotactical}.
For bilayer systems, the inter-layer spacing was fixed at 5 Å, and a 15 Å vacuum layer was set to effectively avoid interactions between periodic images and ensure the independence of the system. 
The WANNIER90 program~\cite{mostofi2014updated,marzari2012maximally} was used to calculate maximally localized Wannier functions to create the low-energy Hamiltonian and the projection matrix. We use cRPA method to compute the effective interaction matrix, where the projector method is adopted to  subdivide the target space from the full Fock space in the Wannier basis. 

We perform calculations on a series of two-dimensional van der Waals materials, extracting their bare interaction $V$, on-site Hubbard parameter $U_o$, and the inter-layer electron repulsion in neighboring layers, denoted $U_{12}\equiv U$. (Note that our definitions of $U_o$ and $U$ differ from the conventional ones.) It should be noted that due  to technical limitations, we were unable to compute $U$  directly  for the Na-adsorbed $\mathrm{NaCr_{2}Cl_{6}}$ system. Instead, we calculated $U$ parameter for $\mathrm{Cr_{2}Cl_{6}}$. As shown in Table~\ref{tab:booktabs}, the $\mathrm{Cr_{2}Cl_{6}}$ bilayer exhibits an inter-layer repulsion $U \sim 0.57$ eV. Since the adsorbed Na atom primarily introduces energy bands near the Fermi level that lie within the target space in cRPA, the difference in $U$ between $\mathrm{Cr_{2}Cl_{6}}$ and $\mathrm{NaCr_{2}Cl_{6}}$ is expected to be small.

Direct cRPA calculations on multi-layer systems are also challenging for $\mathrm{FeP_{35}}$. As an alternative, we computed $U$ for $\mathrm{FeP_{7}}$ bilayers, yielding $U \sim 0.15$ eV (see Table~\ref{tab:booktabs}). Assuming this $U$ value applies to $\mathrm{FeP_{35}}$, it is approximately half the bandwidth $W \approx 0.37$ eV of $\mathrm{FeP_{35}}$.


\end{document}